\documentclass[preprint,12pt]{elsarticle}

\usepackage{amssymb}
\usepackage{amsmath}
\usepackage{mathrsfs}
\usepackage{xcolor}
\usepackage[english]{babel}

\journal{}

\begin{document}

\begin{frontmatter}



\title{Hydrostatic mass of galaxy clusters within some theories of gravity}

 \author[label1]{Feri Apryandi}
 \affiliation[label1]{organization={Study Program of Physics, Universitas Pendidikan Indonesia},
             city={Bandung},
             country={Indonesia}}
 \ead{feri.apryandi@upi.edu}
 \author[label2]{M. Lawrence Pattersons} 
 \affiliation[label2]{organization={Theoretical High Energy Physics Group, Department of Physics, Institut Teknologi Bandung},
             addressline={Jl. Ganesha 10},
             city={Bandung},
             postcode={40132},
             country={Indonesia}}


\begin{abstract}
The mass of galaxy clusters (GCs) can be determined by calculating the hydrostatic equilibrium equation. In this work, we derive the hydrostatic mass of GCs within Eddington-inspired Born-Infeld (EiBI) theory, beyond Horndeski gravity (BHG), and modified emergent Newtonian gravity (MENG) with generalized uncertainty principle (GUP) correction. We apply the formulations on the masses of 10 GCs. We compare our results with the Newtonian mass of GCs. Within a regime, we get an insight that all formulations could match the Newtonian mass. Thus, the impact of the modified theories of gravity used in this work can be neglected in this regime. The noteworthy impact starts if we set $\kappa=5\times10^{40}$ m$^2$ for EiBI theory, $\Upsilon=-0.1655\times10^{69}$ for BHG, and $\beta_0=-1.656\times10^{110}$ for MENG. We also compare our results from EiBI theory and BHG with the baryonic masses $M_{bar}$ of the GCs. A better linear fit is achieved by EiBI theory with $\kappa=5.80\times10^{40}$ m$^2$, which gives the slope $\mathcal{M}$ of $0.126\pm0.086$. This value is closer to unity than the one of BHG. This leads us to the fact that EiBI theory is more effective than BHG in alleviating the mass discrepancy between hydrostatic mass and baryonic mass in GCs. Nevertheless, neither EiBI theory nor BHG completely addresses the mass discrepancy problem.
\end{abstract}



\begin{keyword}
BHG
\sep EiBI theory
\sep emergent gravity
\sep galaxy clusters
\sep hydrostatic mass



\end{keyword}

\end{frontmatter}



\section{Introduction}
\label{sec1}

Galaxy clusters (GCs) are the largest virialized astronomical structures in the universe \cite{Holanda2015}. They are generally defined as self-gravitating systems of galaxies that may be identified in observations as the systems exhibiting one or more of the following features:  a significantly higher concentration of galaxies relative to the average distribution of the large-scale structure of the universe;  massive concentration(s)  of  X-ray emitting gas, often associated with the  Sunyaev-Zel’dovich (SZ) effect on the Cosmic    Microwave    Background    Radiation; and a non-random gravitational lensing shear map of background galaxies \cite{Premadi2021}. As a relatively large structure, GCs are magnificent laboratories for constraining fundamental physics, such as bounding the neutrino mass, graviton mass, or limits on primordial non-Gaussianity. GCs also have an important role in cosmology due to the many discoveries based on optical, X-ray, and microwave surveys \cite{Gupta2020}.

GCs are full of hot, diffuse plasma, with typical temperatures $T \sim 7\times 10^7$ K and electron densities $n_e\sim 10^{-3}$ cm$^{-3}$, This intracluster medium leads to emission of copious X-ray, with typical cluster X-ray luminosities $L_x \sim 10^{43} - 10^{45}$ ergs s$^{-1}$. The total mass of intracluster gas generally transcends the mass of all galaxies in a rich cluster. The gas might be almost hydrostatic. From the distribution of the intracluster plasma, one can obtain the total gravitational mass of the cluster and its distribution \cite{Sarazin1992}.

Traditionally, the mass of GCs can be determined using three methods, i.e. velocity dispersion, X-ray profiles or SZ observations assuming hydrostatic equilibrium, and gravitational lensing. Apart from gravitational lensing, all those methods formulate GC's mass within Newtonian approximation \cite{Gupta2020}. The reviews on all mentioned methods can be found in Refs. \cite{Sarazin1986,Bartelmann1995,Voit2005,Allen2011,Kravtsov2012,Ettori2013,Hoekstra2013,Munari2013,Vikhlinin2014,Li2023}.

Bambi \cite{Bambi2007} profoundly worked on the problem of GC's mass. Starting from an assumption that dark energy (DE) behaves like cosmological constant $\Lambda$ on a relatively small scale, he discussed the effects of a non-zero cosmological constant in the Newtonian limit. As a result, he derived the general relativity (GR) correction for the mass of GCs, which leads to such an effective mass of GCs, i.e. $M_0 -\frac{8}{3}\pi r^3 \rho_\Lambda$, where $M_0$ is the Newtonian mass of GCs, and $\rho_\Lambda$ is the energy density associated with the cosmological constant $\Lambda=8\pi G\rho_\Lambda$, and $G$ denotes the universal gravitational constant. It is worth noting that the masses of GCs obtained from velocity dispersion and X-ray profile methods are the effective masses, while the masses obtained from gravitational lensing are the pure Newtonian masses of GCs. Furthermore, even though gravitational lensing is a GR phenomenon, Bambi argued that the gravitational lensing measurement of GC's mass is not affected by $\Lambda$. Based on Bambi's formulation, Desai \cite{Desai2022} calculated the masses of several GCs. He found that the impact of general relativistic corrections to hydrostatic masses of galaxy clusters using the Kottler metric (i.e. the metric used in Bambi's work \cite{Bambi2007}) is negligible.

Gupta \& Desai \cite{Gupta2020} proposed another formulation to find GR-type correction on GC's mass, i.e. using the Tolman-Oppenheimer-Volkoff (TOV) equation. They then applied the GC's mass formulation they obtained to calculate the hydrostatic masses of twelve GCs. It is important to keep in mind that the TOV equation is the GR form of the Newtonian hydrostatic equilibrium equation. The differences between the TOV-based mass and the Newtonian mass are corresponding to $\mathcal{O}(10^{-5})$, so they can be neglected. It is concluded that for all practical purposes, we can safely use the Newtonian equation of hydrostatic equilibrium for any dynamical analysis with GCs.

Another attempt to obtain the mass of GCs was achieved by Apryandi et. al. \cite{Feri2021}. Using the generalized uncertainty principle (GUP) and following Verlinde's emergent gravity or entropic gravity procedure \cite{Verlinde2011} to get Newtonian gravity formula, they derived the modified emergent Newtonian gravity (MENG). The formalism was then used to derive a formula to calculate the mass of GCs. However, their work has not applied the formula to calculate the masses of GCs, as has been done by Desai \cite{Desai2022} and Gupta \& Desai \cite{Gupta2020}. {Additionally, there are inconsistencies in the derivation of the hydrostatic equation in Ref. \cite{Feri2021}. These issues can be addressed using the derivation presented in Ref. \cite{Belfaqih2021}, which also considers the GUP.}

Basically, Verlinde's theory of emergent gravity describes gravity as an emergent phenomenon rather than a fundamental force. This reasoning can be applied to de Sitter space, suggesting that gravity behaves differently on galaxy and GC scales. Such excess gravity might provide an alternative to dark matter (DM) \cite{Tamosiunas2019}. The emergent gravity paradigm aims to derive and interpret the field equations of gravitational theories within a thermodynamic framework \cite{Padmanabhan2015}. For further discussions on emergent gravity, see Refs. \cite{Verlinde2011,Tamosiunas2019,Padmanabhan2015}.

Notably, Tamosiunas et al. \cite{Tamosiunas2019} found that GR is preferred over emergent gravity across all tested datasets of galaxies and GCs. They discovered that the simultaneous emergent gravity fits of the X-ray and weak lensing datasets are significantly worse than those provided by GR with cold dark matter. For the Coma cluster, the predictions from emergent gravity and GR agree in the range of 250$-$700 kpc. In contrast, at around 1 Mpc scale, emergent gravity total mass predictions are larger by a factor of 2. For the cluster stack the predictions are only in good agreement at around the 1$-$2 Mpc scales, while for the radius $r \gtrsim$ 10 Mpc, emergent gravity is in strong tension with the data. It is important to keep in mind that GUP's effect is not included in the work of Tamosiunas et al. Although controversial \cite{Kobakhidze2011,Visser2011,Dai2017}, Verlinde's theory of emergent gravity has good agreement with some observations (please see Refs. \cite{Brouwer2017,Lelli2017,Tortora2018,Yoon2024}).

Other than the work of Apryandi et al. \cite{Feri2021}, some attempts to involve GUP in Verlinde's theory have been achieved. Feng et al. \cite{Feng2016} found that the GUP-corrected entropic force is related to the properties of the black holes and the Planck length. Furthermore, based on the GUP corrected entropic force, they derive the modified Einstein equations and the modified Friedmann equation. Kibaroğlu \cite{Kibaroğlu2019} obtained generalized Einstein equations with cosmological constant by including GUP in emergent gravity and using the modified Unruh temperature. On the other side, motivated by the existence of some white dwarfs that are more massive than the Chandrasekhar mass limit, Belfaqih et al. \cite{Belfaqih2021} included GUP in Verlinde's theory of emergent gravity to find mass and radius relation of white dwarfs. Jizba et al. \cite{Jizba2023} unified GUP and Tsallis thermostatistics, and then applied their achievement to Verlinde's theory. A discussion on Tsallis thermostatistics can be found in Ref. \cite{Naudts2011}.

On the other hand, beyond-GR theories or modified theories of gravity are also important to be considered in the studies of objects in the cosmos. For decades, people have been witnessing the evidence that, if gravity is described by GR, there should be a considerable amount of DM in galaxies and GCs. Furthermore, DE has become a problem that needs to be explained as the expansion of the Universe is still accelerated. However, if GR is correct, it must be around 96 \% of the Universe in the form of energy densities that do not have any electromagnetic interaction \cite{Clifton2012}. In the context of GCs, observations of GCs suggest that they contain fewer baryons (gas plus stars) than the cosmic baryon fraction. This so-called “missing baryon” is surprising for the most massive GCs, which are expected to be representative of the cosmic matter content of the Universe (baryons and DM) \cite{Rasheed2011}. These facts have led us to a speculation on the possibility that GR may not, apparently, be the correct theory of gravity to describe the Universe on the largest scales. Our dark universe may be just another sign that we have to go beyond Einstein’s GR \cite{Clifton2012}.

An example of a modified gravity theory is the Eddington-inspired Born-Infeld (EiBI) theory. Santos \& Santos \cite{Santos2015} introduced the concept of geometrical mass for GCs within the framework of EiBI theory. They highlighted the primary issue of mass discrepancy in GCs, which is typically addressed by postulating the existence of cold, pressureless DM. However, the nature of DM remains unknown, and the only evidence for its existence so far is gravitational. Thus, another possible way to overcome the problem of discrepancy of the GC's mass is by modifying the theory of gravity. In their work, they derived the virial theorem in EiBI theory and showed that the geometrical mass, which is a mathematical artifact in the theorem, may account for the virial mass discrepancy in GCs.

Another example of the modified theory of gravity is beyond Horndeski gravity (BHG), which is a modified gravity containing the screening mechanism if $R << r_v$, where $R$ is the radius of the corresponding astrophysical body, and $r_v$ is the Vainshtein radius. Here $r_v$ is defined as the transition between the screened and unscreened regimes inside astrophysical bodies \cite{Rosyadi2019}. The Vainshtein mechanism is discussed in Refs. \cite{Jain2010,Koyama2015} and the references therein. Sakstein et. al. \cite{Sakstein2016} used GCs to obtain new constraints on BHG. In their work, they presented the so-called thermal mass, which was previously introduced by Terukina et.al. \cite{Terukina2014}. In Ref. \cite{Sakstein2016}, the thermal mass is equivalent to the weak-lensing mass of GCs, and found by solving the hydrostatic equilibrium equation for the thermal part i.e. the part deduced from X-ray temperature profiles. 

Motivated by the previous studies, in this work, {we derive the formulation for the hydrostatic mass of GCs within EiBI theory, BHG, and MENG. We then calculate the hydrostatic mass of several GCs as presented in Ref. \cite{Gupta2020}, allowing free parameters which yield negligible corrections. Additionally, we vary the free parameters of the EiBI theory and BHG beyond the values that yield negligible corrections compared to the Newtonian mass of GCs. The resulting values from these varied parameters are subsequently compared with the baryonic mass $M_{bar}$ of the GCs, as provided in Ref. \cite{Rahvar2014}. It should be noted that the MENG results are not compared to $M_{bar}$, as the quantum corrections from the GUP are expected to have minimal impact. We anticipate that the baryonic mass and hydrostatic mass will essentially agree if the free parameters of the EiBI theory and BHG are correctly represented by our formulation.}

This paper is organized as follows. In section 2, we explain the construction of formalism we use, where the derivations of hydrostatic mass of GCs within EiBI theory and BHG are presented. {Moreover, the mass of GCs within MENG formulation derived in Ref. \cite{Feri2021} is re-derived by using the formalism in Ref. \cite{Belfaqih2021}}. In section 3, we present and discuss the numerical results of GC masses calculation. Lastly, section 4 contains the concluding remarks of this work.

\section{Formalisms}
\label{sec2}

To make the discussion self-contained, {subsection 2.1 derives the hydrostatic mass of GCs within EiBI theory; subsection 2.2 presents the derivation within BHG; and subsection 2.3 provides the re-derivation within MENG.}
\subsection{Hydrostatic mass of GCs within EiBI theory}
\label{subsec2}
Firstly, we look at the non-relativistic limit of EiBI theory which is shown by the following Poisson equation \cite{Banados2010,Wibisono2018}
\begin{eqnarray}
    \nabla^2\Phi=4\pi G\rho+\frac{\kappa}{4}G\:\nabla^2\rho.
\end{eqnarray}

Here $\Phi$ denotes gravitational potential, while $\kappa$ denotes the free parameter of EiBI theory in the unit of m$^2$. The equation of hydrostatic equilibrium within EiBI theory is given by
\begin{eqnarray}
	\frac{dP}{dr}=-\frac{GM\rho}{r^{2}}-\frac{\kappa}{4}\:G\rho\:\frac{d\rho}{dr},\label{dpdreibi}
	\end{eqnarray}
where $P$ denotes pressure, $G$ is the universal gravitational constant, and $\rho$ denotes mass density.

{Here we use the assumption that the Equation of State (EOS) of GCs is satisfied by the EOS of ideal gases. Such EOS was also used by Gupta \& Desai \cite{Gupta2020}. For the ideal gas,
\begin{equation}
    \frac{dP}{dr}=\frac{k_b}{\mu\:m_p}\left(T\:\frac{d\rho}{dr}+\rho\:\frac{dT}{dr}\right),\label{dpdridealgas]}
\end{equation}
where $k_b$ denotes Boltzmann's constant, $\mu$ is the mean molecular weight of the cluster, $m_p$ is the mass of the proton, and $T$ denotes the absolute temperature.}

By combining Eq. (\ref{dpdreibi}) and Eq. (\ref{dpdridealgas]}), it is easy to obtain the mass of GCs within EiBI theory
 \begin{eqnarray}
      M_{EiBI}(r)=-\frac{rk_{b}}{\mu m_{p}G\rho}T\rho\left(\frac{r}{\rho}\frac{d\rho}{dr}+ \frac{r}{T}\frac{dT}{dr}\right)-r^2\:\frac{\kappa}{4}\frac{d\rho}{dr}.
 \end{eqnarray}
 Considering that $\left(\frac{r}{\rho}\frac{d\rho}{dr}+ \frac{r}{T}\frac{dT}{dr}\right)=\left(\frac{d \ln \rho}{d \ln r} + \frac{d \ln T}{d \ln r}\right)$, we obtain
\begin{eqnarray}
M_{EiBI}(r)= -\frac{rk_{B}}{\mu m_{p}G}\:T\left(\frac{d \ln \rho}{d \ln r} + \frac{d \ln T}{d \ln r}\right)-\frac{r^2\kappa}{4}\frac{d\rho}{dr}.    
\end{eqnarray}
By substituting the values of the constants, we finally find the mass of GCs within EiBI theory in the unit of $\mathscr{M}_\odot\equiv10^{14} \:M_\odot$, i.e.
\begin{eqnarray}
    M_{EiBI}(r)=-\left[3,7 \times10^{-4}\:T\:r\left(\frac{d \ln \rho}{d \ln r} + \frac{d \ln T}{d \ln r}\right)+\frac{r\kappa \rho}{4{\mathscr{M}_\odot}}\left(\frac{d \ln \rho}{d \ln r}\right)\right]\:\mathscr{M}_\odot.\label{meibi}
\end{eqnarray}

\subsection{Hydrostatic mass of GCs within BHG}
\label{subsec3}
The weak-field limit of BHG is shown by the following equation \cite{Rosyadi2019}
\begin{equation}
    \nabla^2\Phi=4\pi G\rho+\frac{\Upsilon}{4}G\frac{d^3M}{dr^3},\label{weakfieldbhg}
\end{equation}
where $\Upsilon$ is a dimensionless BHG parameter that characterizes the strength of the gravity modifications. It is worth noting that due to a higher concentration of the mass towards the center, the second term of the right-handed side of Eq. (\ref{weakfieldbhg}) corresponds to a weakening of gravity if $\Upsilon > 0$ and a strengthening when the converse is true \cite{Sakstein2015}.

The hydrostatic equilibrium equation within BHG reads \cite{Sakstein2015,Saito2015}
\begin{equation}
    \frac{dP}{dr}=-\frac{GM\rho}{r^{2}}-\frac{\Upsilon G\rho}{4}\frac{d^{2}M}{dr^{2}},\label{dpdridealbhg}
\end{equation}

By combining Eq. (\ref{dpdridealbhg}) and the equation of hydrostatic equilibrium of ideal gas shown by Eq. (\ref{dpdridealgas]}), one can easily find the hydrostatic mass of GCs within BHG
 \begin{equation}
     M_{BHG}=-\frac{rk_{b}T}{\mu m_{p}G}\left(\frac{d \ln \rho}{d \ln r} + \frac{d \ln T}{d \ln r}\right)-\frac{\Upsilon r^2}{4}\frac{d^2M}{dr^2}\label{diffeqbhg}.
 \end{equation}
To solve $\frac{d^2M}{dr^2}$, we use the fact that
\begin{eqnarray}
\frac{dM}{dr}= 4 \pi r^2 \rho.\label{dmdrbhg}
\end{eqnarray}
The second derivative of $M$ with respect to $r$ gives
\begin{eqnarray}
\frac{d^2M}{dr^2}= 4 \pi r^2 \frac{d\rho}{dr}+8 \pi r \rho.
\end{eqnarray}
With a little calculation, one can find
\begin{eqnarray}
\frac{M_{BHG}(r)}{\mathscr{M}_\odot} =-\frac{rk_{b}T}{\mu m_{p}G}\left(\frac{d \ln \rho}{d \ln r} + \frac{d \ln T}{d \ln r}\right) -\frac{\Upsilon \pi r^3 \rho}{{\mathscr{M}_\odot}}\left(2+\frac{d \ln\rho}{d \ln r}\right).
\end{eqnarray}
By substituting the values of the constants, we finally find the mass of GCs within BHG in the unit of ${\mathscr{M}_\odot}$, i.e. 
\begin{eqnarray}
M_{BHG} =\left[{-3,7 \times10^{-4}}\:T\:r\left(\frac{d \ln \rho}{d \ln r} + \frac{d \ln T}{d \ln r}\right)-\frac{\Upsilon \pi r^3 \rho}{{\mathscr{M}_\odot}}\left(2+\frac{d \ln\rho}{d \ln r}\right)\right]\mathscr{M}_\odot.\label{mbhg}
\end{eqnarray}

\subsection{Hydrostatic mass of GCs within MENG}
\label{subsec1}

In this subsection, {we derive the hydrostatic mass of GCs within MENG. The derivation of hydrostatic equilibrium equation with GUP correction is based on the one presented in Ref. \cite{Belfaqih2021}.} It is worth keeping in mind the ordinary unmodified Heisenberg uncertainty, i.e. $\Delta x \Delta p \geq \frac{\hbar}{2}$. Here we use the quadratic GUP, which reads
\begin{equation}
    \Delta x\Delta p\geq \frac{\hbar}{2}\left(1+\beta (\Delta p)^2\right),\label{QGUP}
\end{equation}
where $\beta=\frac{\beta_0 l_p^2}{\hbar^2}$, with $\beta_0$ is a free dimensionless parameter of GUP, and Planck length $l_p=\sqrt{\frac{G\hbar}{c^3}}\approx$ 10$^{-35}$ m. {The inequality shown by Eq. (\ref{QGUP}) would give}
\begin{equation}
    {\Delta x \geq \hbar \sqrt{\beta}.}
\end{equation}

{It has to be noted that Verlinde's proposal is based on Bekenstein's formula \cite{Bekenstein1973} that relates the entropy and the surface area of black holes. The increase of the black hole's area when absorbing a classical particle whose energy $E$ and size $\Delta x$ reads}
\begin{eqnarray}
    {\Delta A\geq\frac{8\pi l_p^2}{\hbar c}E\Delta x\geq\frac{8\pi l_p^2}{\hbar}\Delta p \Delta x}. \label{deltaA}
\end{eqnarray}

{Solving Eq. (\ref{QGUP}) to first order in $\beta$ and then substituting the solution into Eq. (\ref{deltaA}) yields}
\begin{eqnarray}
    {\Delta A\geq4\pi l_p^2\left[1+\frac{\hbar^2\beta}{4(\Delta x)^2}\right].}
\end{eqnarray}
{Ref. \cite{Belfaqih2021} agreed with Ref. \cite{Scardigli2015} that $\Delta x$ can be estimated as}
\begin{eqnarray}
    {\Delta x \simeq \frac{4\pi GM}{c^2}.}
\end{eqnarray}
{With this estimation, the first-order lower bound for the area increase in terms of $\beta$ is obtained as}
\begin{eqnarray}
    {\Delta A_{min}\simeq4\pi l_p^2\lambda\left[1+\frac{\hbar^2\beta}{4\pi A}\right],}
\end{eqnarray}
{where $A=4\pi \left(\frac{2GM}{c^2}\right)^2$ denotes the area of the black hole's event horizon, and $\lambda$ is a constant that will be fixed later. For entropy, the minimum increment is typically taken as one bit, such that $\Delta S_{min}=b=\ln 2$. Thus, from these results, we can write}
\begin{eqnarray}
    {\frac{dS}{dA}\simeq\frac{\Delta S_{min}}{\Delta A_{min}}=\frac{b}{4\pi\lambda l_p^2\left[1+\frac{h^2\beta}{4\pi A}\right]}.}
\end{eqnarray}
{Therefore,}
\begin{eqnarray}
    {\frac{dS}{dA}\simeq\frac{b}{4\lambda\pi l_p^2}\left[1-\frac{h^2\beta}{4\pi A}\right].}\label{dSdA}
\end{eqnarray}
{Integrating Eq. (\ref{dSdA}) would give}
\begin{equation}
    {S(A,\beta)=\frac{b}{4\lambda\pi l_p^2}\left[A-\frac{\hbar^2\beta}{4\pi}\ln\left(\frac{A}{4l_p^2}\right)\right].}\label{entropy}
\end{equation}
{If we take a limit $\beta\rightarrow 0$, Eq. (\ref{entropy}) is reduced to the standard Bekenstein-Hawking entropy $S_{BH}$. It can be identified that}
\begin{equation}
    {\frac{b}{4\pi\lambda l_p^2}=\frac{k_b c^3}{4 G \hbar}.}
\end{equation}
{Eq. (\ref{entropy}) now writes}
\begin{equation}
    {S(A, \beta)=\frac{k_b c^3}{4G\hbar}\left[A-\frac{\hbar^2\beta}{4\pi}\ln\left(\frac{A}{4 l_p^2}\right)\right].\label{modifentropy}}
\end{equation}

{Please note that Verlinde's idea relies on the holographic principle which states that microscopic informations (i.e. the bits) of a closed surface are embedded on the surface on radius $r$. The amount of bits is given by}
\begin{eqnarray}
    {N=\frac{4S}{k_b}}.\label{n}
\end{eqnarray}
{In Ref. \cite{Verlinde2011}, Verlinde conjectured that the bits carried the energy of the system and the energy is divided evenly over the bits. The equipartition theorem gives the temperature}
\begin{eqnarray}
    {T=\frac{2E}{N\:k_b},}
\end{eqnarray}
{where $E=Mc^2$ denotes the total energy of the system}.

{Following Bekenstein's argument, as a test particle of mass $m$ approaches the surface by an amount $\Delta x$, the corresponding change in entropy associated with the boundary information reads}
\begin{eqnarray}
    {\Delta S=2\pi k_b\frac{\Delta x}{\hbar}mc.}
\end{eqnarray}

{Recall the relation of $\Delta S$ to entropic force}
\begin{eqnarray}
    {F\Delta x=T\Delta S,}\\
    {F=\frac{4\pi Mmc^3}{\hbar N}.}\label{f}
\end{eqnarray}

{On the other side, by substituting Eq. (\ref{modifentropy}) into Eq. (\ref{n}), we can obtain}
\begin{eqnarray}
    {N=\frac{c^3}{G\hbar}A\left[1-\frac{\hbar^2\beta}{4\pi A}\ln\left(\frac{A}{4l_p^2}\right)\right].}\label{modifn}
\end{eqnarray}
{Substituting Eq. (\ref{modifn}) into Eq. (\ref{f}) gives the modified Newtonian force}
\begin{eqnarray}
    {\mathcal{F}=G\frac{Mm}{r^2}\left[1+\frac{\hbar^2\beta}{16\pi^2 r^2}\ln\left(\frac{\pi r^2}{l_p^2}\right)\right].}
\end{eqnarray}

{Now consider a spherical system whose mass $M(r)$. The unmodified Newtonian force $dF$ acted on a spherical shell with thickness $dr$ and surface area $dA$ due to the matter inside it writes}
\begin{equation}
    {dF=A\:dP=-G\frac{M(r)\:dM(r)}{r^2},}
\end{equation}
{where $dM(r)$ denotes the mass of the spherical shell, and $dP$ is the outward pressure balancing the gravitational attraction. Recall $dM(r)=\rho(r)A\:dr$, so}
\begin{equation}
    {\frac{1}{\rho(r)}\frac{dP}{dr}=-\frac{GM}{r^2}}.
\end{equation}
We can easily obtain
\begin{equation}
    {\frac{dP}{dr}=-\frac{GM(r)\rho(r)}{r^2}}\label{hydrounmodif}.
\end{equation}

{On the other side, the Newtonian force within MENG reads}
\begin{eqnarray}
    {d\mathcal{F}=A\:dP=-G\frac{M(r)dM(r)}{r^2}\left[1+\frac{\hbar^2\beta}{16\pi^2 r^2}\ln\left(\frac{\pi r^2}{l_p^2}\right)\right]}.
\end{eqnarray}
{For this case, Eq. (\ref{hydrounmodif}) now becomes}
\begin{equation}
    {\frac{dP}{dr}=-\frac{GM(r)\rho(r)}{r^2}\left[1+\frac{\hbar^2\beta}{16\pi^2 r^2}\ln\left(\frac{\pi r^2}{l_p^2}\right)\right]}.\label{dpdrmeng}
\end{equation}
{By combining Eq. (\ref{dpdrmeng}) and Eq. (\ref{dpdridealgas]}), one can obtain the hydrostatic mass of GCs within MENG}
\begin{eqnarray}
    {M_{MENG}=-\frac{k_b\:T\:r}{G\:\mu\: m_p}\left(\frac{d\ln\rho}{d\ln r}+\frac{d\ln T}{d\ln r}\right)\left[1+\frac{\hbar^2\beta}{16\pi^2 r^2}\ln\left(\frac{\pi r^2}{l_p^2}\right)\right]^{-1}.\label{mmeng}}
\end{eqnarray}
\begin{figure}
    \centering
    \includegraphics[width=11cm]{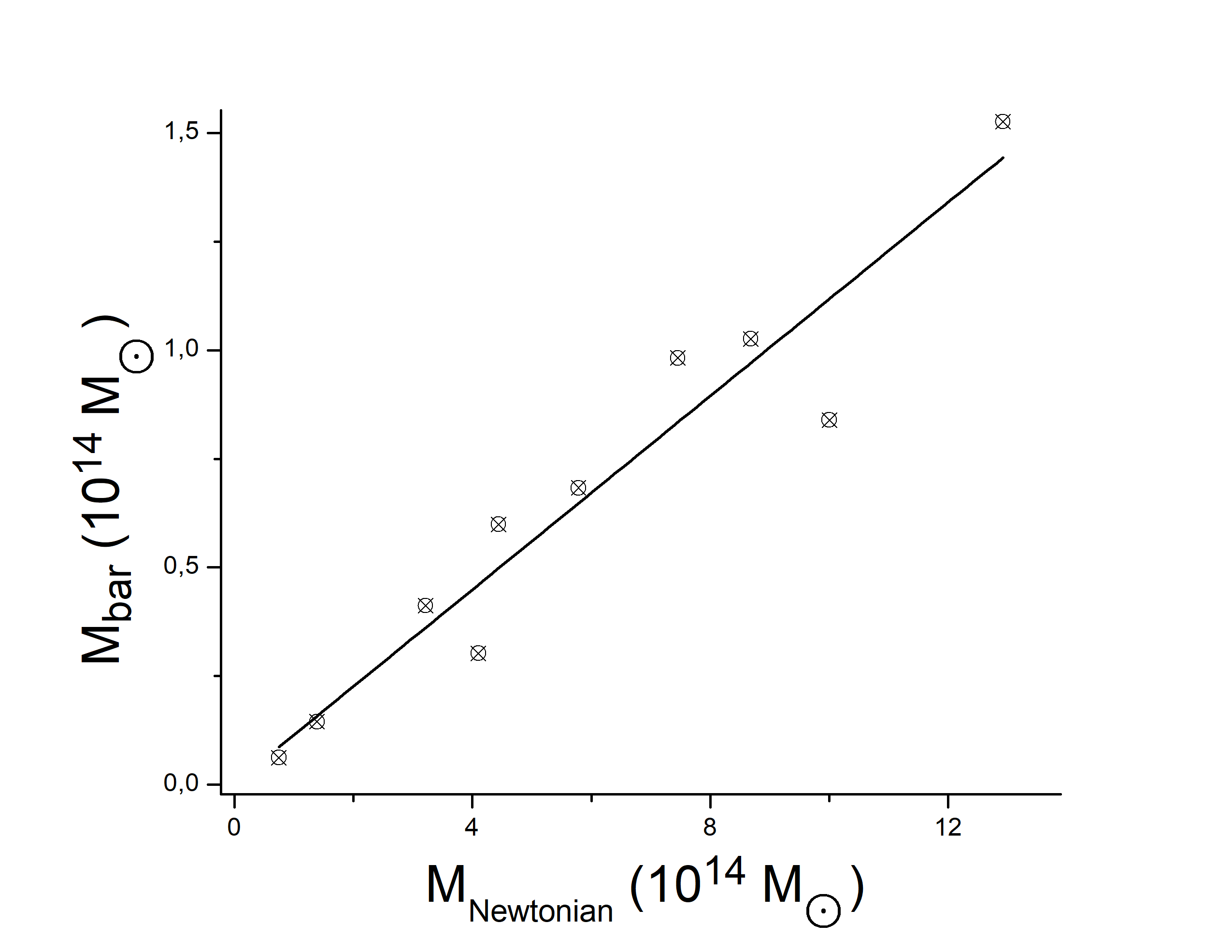}
    \caption{The linear fit of the relation between Newtonian hydrostatic masses and the observed baryonic mass $M_{bar}$ of 10 GCs.}
    \label{fig:Newtonianfitting}
\end{figure}

\section{Numerical results}
Vikhlinin et. al. \cite {Vikhlinin2006} have obtained the density and temperature profiles for a total of 13 nearby relaxed galaxy clusters from Chandra and ROSAT observations. These profiles are valid up to approximately 1 Mpc \cite{Gupta2020}. We use the data of 10 GCs provided in Ref. \cite{Vikhlinin2006} to calculate the Newtonian mass of GCs. To numerically calculate the hydrostatic mass of GCs according to MENG, EiBI theory, and BHG, we use Eq. (\ref{mmeng}), Eq. (\ref{meibi}), and Eq. (\ref{mbhg}), respectively.

\begin{figure}[htp]
    \centering
    \includegraphics[width=11cm]{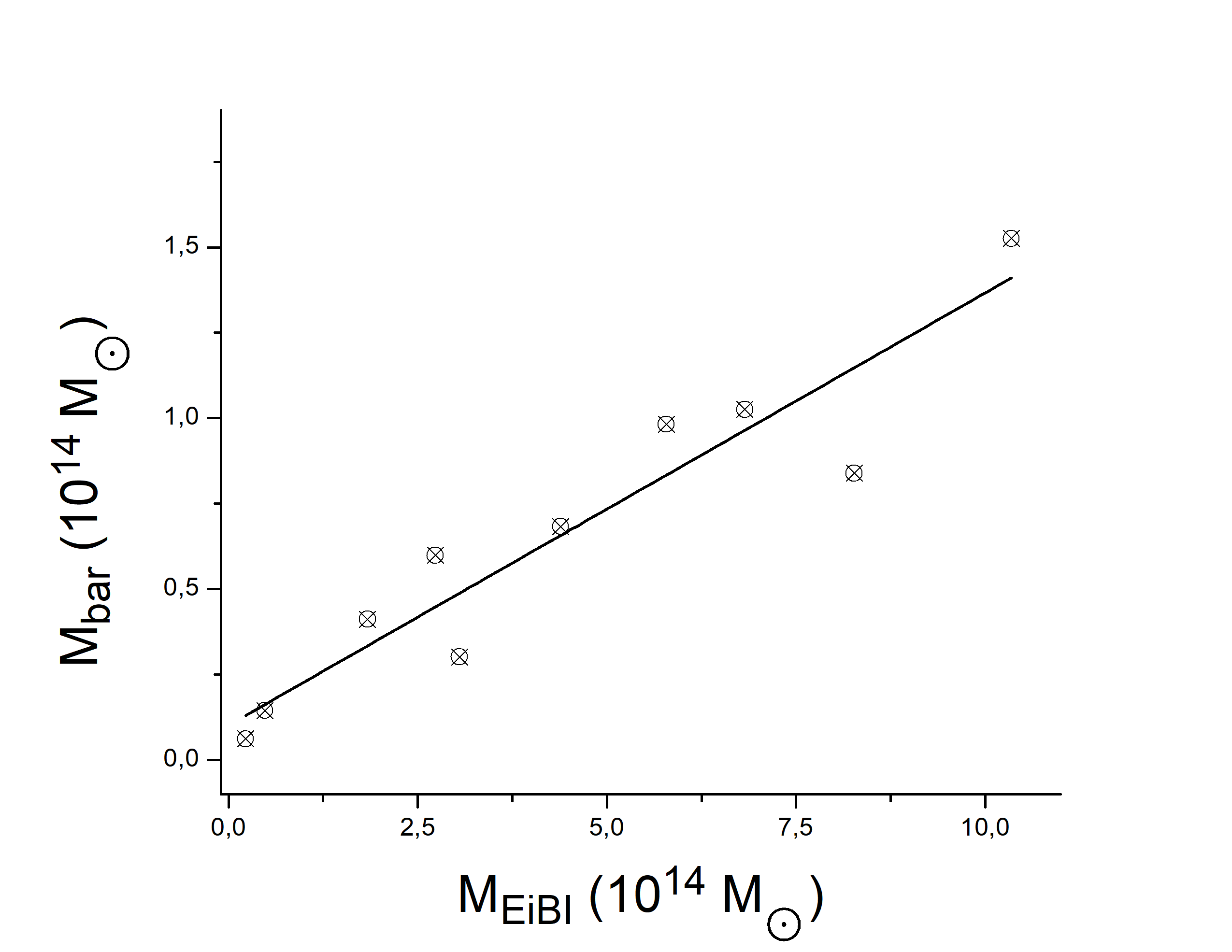}\\
~~~~~~~~~~~~~~~~~~~~~~~~~~~~\textbf{(a)}~~~~~~~~~~~~~~~~~~~~~~~~~\\
    \includegraphics[width=11cm]{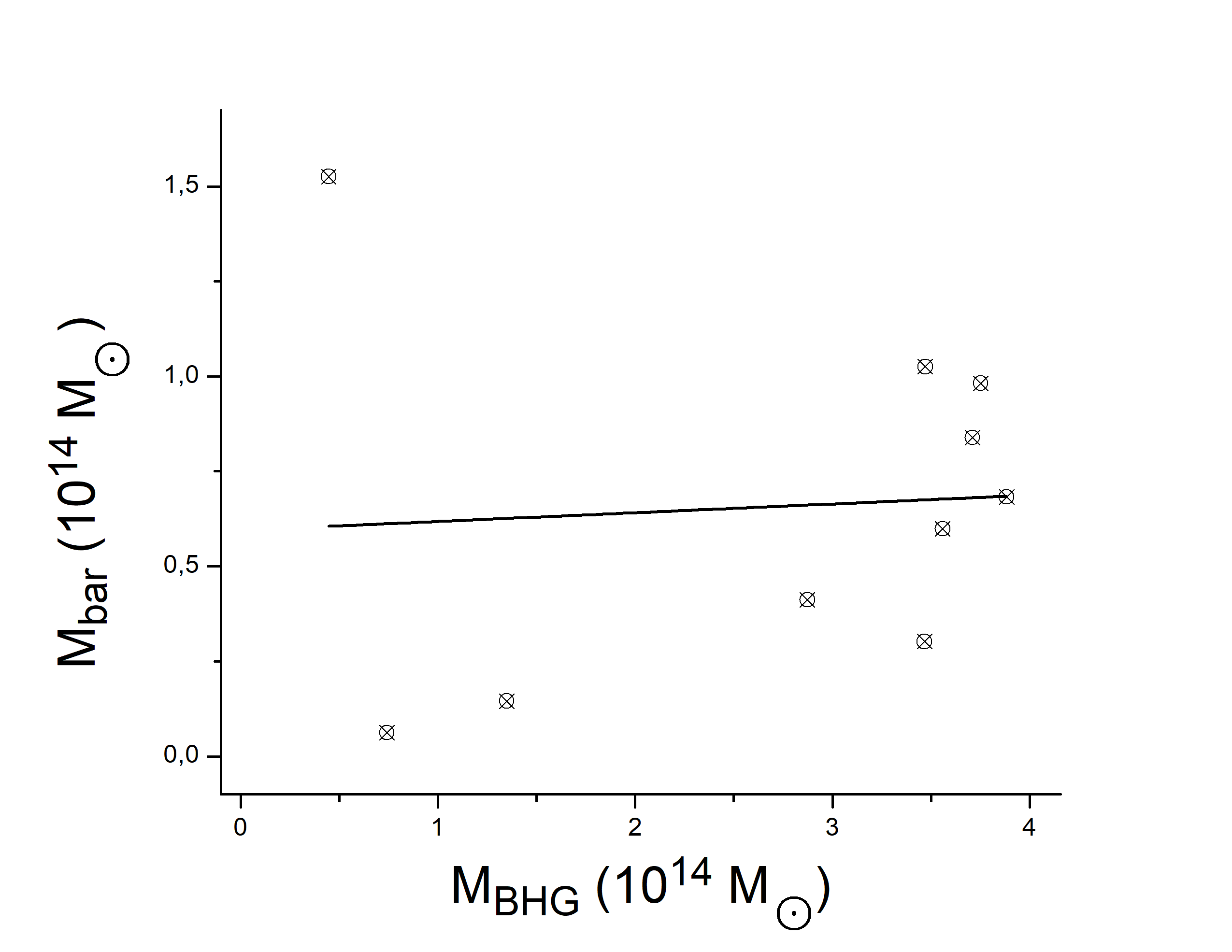}\\
~~~~~~~~~~~~~~~~~~~~~~~~~~~~\textbf{(b)}~~~~~~~~~~~~~~~~~~~~~~~~~
    \caption{{The linear fit of the relation between hydrostatic masses and the observed baryonic mass $M_{bar}$ of 10 GCs. (a) $M_{EiBI}$ with $\kappa=5.80\times10^{40}$ m$^2$. (b) $M_{BHG}$ with $\Upsilon=1.65\times10^{71}$.}}
    \label{fig:pic}
\end{figure}

\begin{table}[]
    \centering
    \begin{center}
    \caption{Fractional difference between
hydrostatic masses of GCs within various theories of gravity and Newtonian mass at $r_{500}$ for 10 clusters. The data of GCs' mass $r_{500}$ are obtained from Ref. \cite{Gupta2020}.}
    \begin{tabular}{ | c | c | c | c | c | c |}
    \hline
    GC & $r_{500}$ & $M_{Newtonian}$ & $\lvert\frac{\Delta M_{MENG}}{M_{Newtonian}}\rvert$ (\%) & $\lvert\frac{\Delta M_{EiBI}}{M_{Newtonian}}\rvert$ (\%) & $\lvert\frac{\Delta M_{BHG}}{M_{Newtonian}}\rvert$ (\%) \\
     & (kpc) & ($\mathscr{M}_\odot$) & \tiny{$\beta_0=-1.656\times10^{110}$} & \tiny{$\kappa=5\times10^{40}$ m$^2$} & \tiny{$\Upsilon=-0.1655 \times 10^{69}$} \\ \hline
    A133 & 1007 $\pm$ 41 & 4.1 & 0.34 & 0.16 & 0.16 \\ 
    A383 & 944 $\pm$ 32 & 3.2 & 0.83 & 0.14 & 0.11\\
    A478 & 1337 $\pm$ 58 & 7.4 & 0.89 & 0.53 & 0.50\\ 
    A907 & 1096 $\pm$ 30 & 4.4 & 1.18 & 0.61 & 0.20\\ 
    A1413 & 1299 $\pm$ 43 & 9.9 & 1.19 & 0.87 & 0.63\\
    A1795 & 1235 $\pm$ 36 & 5.8 & 0.18 & 0.21 & 0.33\\ 
    A1991 & 732 $\pm$ 33 & 1.4 & 0.56 & 0.53 & 0.03\\ 
    A2029 & 1362 $\pm$ 43 & 8.6 & 1.08 & 0.75 & 0.60\\
    A2390 & 1416 $\pm$ 48 & 13 & 0.14 & 0.17 & 0.97\\ 
    MKW 4 & 634 $\pm$ 28 & 0.75 & 0.71 & 0.59 & 0.01\\
    \hline
    \end{tabular}\\
\end{center}
    \label{tab1}
\end{table}

We find that for MENG, the noteworthy impact starts if we set $\beta_0=-1.656\times10^{110}$; $\kappa=5\times10^{40}$ m$^2$ for EiBI theory, and $\Upsilon=-0.1655\times10^{69}$ for BHG, as shown in Table 1. Even so, the impacts which are shown by the mass corrections ($\Delta M_{MENG}$, $\Delta M_{EiBI}$, and $\Delta M_{BHG}$) are so small compared to the uncertainties of Newtonian masses of GCs in Ref. \cite{Vikhlinin2006}, which are about 10\%. It may be surprising if we compare our results with the results obtained in Refs. \cite{Belfaqih2021}, \cite{Wibisono2018}, and \cite{Rosyadi2019}. In Ref. \cite{Belfaqih2021}, $\beta_0$ could give a significant impact in the order of $10^{43}$; in Ref. \cite{Wibisono2018}, $\kappa=0.2\times10^{9}$ m$^2$ could result in a meaningful impact; while in Ref. \cite{Rosyadi2019}, $\Upsilon=0.2$ also give a noteworthy impact. It is not surprising at all if we realize that the objects used are white dwarf in Ref. \cite{Belfaqih2021}, polytropic star in Ref. \cite{Wibisono2018}, and brown dwarf in Ref. \cite{Rosyadi2019}. Compared to GCs, those objects have different structures and EOS. So it is clear that GCs have their own values of the free parameters (i.e. $\beta_0$, $\kappa$, and $\Upsilon$) to conduct some impact on the mass. However, although $\beta_0$ is a free parameter, setting its value to produce a significant impact observable on astronomical scales is not appropriate, as $\beta_0$ originates from quantum corrections.

{It is important to emphasize that the choice of the sign of free parameters in Table 1 is not strictly governed by any specific rule. The primary purpose of Table 1 is to identify the parameter values, particularly those farther from zero, at which their impact on the mass within modified gravity theories becomes more pronounced. Moreover, since $\kappa$, $\Upsilon$, and $\beta_0$ are free parameters, it is evident from Refs. \cite{Belfaqih2021, Rosyadi2019, Wibisono2018} that these parameters can take either positive or negative values.}

Our result leads to an argument that if the values of the free parameters $\beta_0$, $\kappa$, and $\Upsilon$ are closer to zero than the values shown in Table 1, one can safely use the Newtonian calculation for hydrostatic mass of GCs, and the corrections of the gravitational theories used in this paper are negligible. The typical result is also obtained within the formalism which includes GR corrections, see Refs. \cite{Gupta2020,Desai2022}.

On the other hand, we also compare the Newtonian mass of GCs and our results with the baryonic mass of GCs shown in Ref. \cite{Rahvar2014}. The Newtonian mass, derived from the hydrostatic equilibrium equation, corresponds to the Tolman-Oppenheimer-Volkoff (TOV) equation in GR. The GC mass obtained from the TOV equation has been studied in Ref. \cite{Gupta2020}, where numerical results indicate that the GR correction is negligible. Therefore, the Newtonian GC mass can be considered equivalent to the GR mass in the Newtonian limit. Furthermore, if the parameter sets of the EiBI theory, BHG, and MENG are set to zero, the formalisms consequently recover to the Newtonian mass. Now consider
\begin{equation}
    \mathcal{M}=\frac{M_{modified\:gravity}}{M_{bar}},
\end{equation}
which represents the slope, with $M_{modified\:gravity}$ could be $M_{MENG}$, $M_{EiBI}$, or $M_{BHG}$. We expect $\mathcal{M}$ should be fundamentally unity. According to Ref. \cite{Rahvar2014}, significant observational uncertainties exist in the determinations of $M_{bar}$. This uncertainty also could exist in the determination of $M_{modified\:gravity}$.

The linear fit of the Newtonian mass shown in Fig. \ref{fig:Newtonianfitting} gives the slope of $0.112\pm0.012$. The linear fit of the EiBI theory with $\kappa=5.80\times10^{40}$ m$^2$ gives the slope of $0.126\pm0.086$, while the BHG with $\Upsilon=1.65\times10^{71}$ gives the slope of $0.023\pm0.119$. The linear fit of EiBI theory is shown in Fig. \ref{fig:pic}(a), while the linear fit of BHG is shown in Fig. \ref{fig:pic}(b). {The slope of the results from EiBI theory is closer to unity compared to that of BHG, indicating that EiBI theory is more effective than BHG in alleviating the mass discrepancy between hydrostatic mass and baryonic mass in GCs, under the same assumption used in Ref. \cite{Rahvar2014} of no dark matter. Moreover, the slope of the results from the EiBI theory is closer to unity compared to those from the Newtonian mass. Nevertheless, the standard error of the linear fit in the EiBI theory is larger than that of the Newtonian mass. However, these slope values remain significantly different from unity, suggesting that neither EiBI theory nor BHG fully resolves the mass discrepancy. The exploration of alternative modified gravity theories with appropriate parameters to address the mass discrepancy problem is deferred to future work.}

It is important to note that for both EiBI theory and BHG when we set $\kappa$ and $\Upsilon$ as positive numbers, the mass of GCs decreases. {As both $\kappa$ and $\Upsilon$ are magnified with negative signs, the hydrostatic mass of GCs increases. The parameter signs used in Fig. \ref{fig:pic}(a) and Fig. \ref{fig:pic}(b) are intended to reduce the mass, bringing it into alignment with the baryonic mass order, specifically around $10^{13} M_\odot$ (see Ref. \cite{Rahvar2014}). As shown in all presented figures, most $M_{bar}$ values fall within this range. These values are notably lower than the mass orders predicted by Newtonian gravity and modified gravity theories used in this work, i.e. EiBI theory and BHG, which are typically around $10^{14} M_\odot$. The parameter 
$\kappa$ is consistently chosen to be positive, as this positive value reduces the GCs' masses compared to those obtained using Newtonian gravity.}

{Similarly, the purpose of selecting a positive value for $\Upsilon$ in the BHG theory, which differs from the notation in the table (as the table uses a negative value), is to decrease the mass values, thereby achieving the baryonic mass order. The value $\Upsilon=1.65\times10^{71}$ sufficiently reduces the masses of some GCs to the baryonic mass order. However, the resulting linear fit slope remains below the slope of Newtonian linear fit. This can be understood by examining Fig. \ref{fig:pic}(b). While some of the GCs' masses in the BHG framework fall within the order of the baryonic mass, their distribution does not display a linear pattern as seen in the cases of $M_{Newtonian}$ shown in Fig. \ref{fig:Newtonianfitting} and $M_{EiBI}$ shown in Fig. \ref{fig:pic}(a). Consequently, the slope of the linear fit for the ratio $M_{BHG}/M_{bar}$ has the lowest value among the cases considered. These findings indicate that even the BHG theory is less effective than Newtonian gravity in addressing the mass discrepancy issue in GCs.}

{Based on the mathematical fact that $M_{EiBI}$ and $M_{BHG}$ decrease when $\kappa$ and $\Upsilon$ are assigned negative values, a trivial thought might arise suggesting that magnifying the negative values of $\kappa$ and $\Upsilon$ could yield better linear fits.} But in fact, it will not happen, since if we set $\kappa$ and $\Upsilon$ exceed the values used in the linear fits shown by Fig. \ref{fig:pic}(a) and Fig. \ref{fig:pic}(b), some masses of GCs reach negative values, which is not physical.

\section{Concluding remarks}
In this work, we have revisited the derivation of MENG with GUP correction derived in Ref. {\cite{Belfaqih2021}, and derive the hydrostatic mass of GCs within MENG with GUP correction.} We also derive the hydrostatic mass of GCs within EiBI theory and BHG. We applied all formulations that we have revisited and derived on 10 GCs and compared the results with the Newtonian mass obtained from Ref. \cite{Gupta2020}. The impact of the theories of gravity used in this work starts if we set $\beta_0=-1.656\times10^{110}$ for MENG, $\kappa=5\times10^{40}$ m$^2$ for EiBI theory, and $\Upsilon=-0.1655\times10^{69}$ for BHG. If the values of the free parameters are set closer to zero than the values mentioned before, the impact is negligible, which means that one can safely use the Newtonian formulation to calculate the hydrostatic mass of GCs.

We compare our results of EiBI theory and BHG with $M_{bar}$. A better linear fit is achieved by EiBI theory with $\kappa_0=5.80\times10^{40}$ m$^2$, which gives the slope $\mathcal{M}$ of $0.126\pm0.086$. This value is closer to unity than the one of BHG. Moreover, the slope of the results from the EiBI theory is closer to unity compared to those from the Newtonian mass. However, the standard error of the linear fit in the EiBI theory is larger than that of the Newtonian mass. This indicates that, compared to BHG, EiBI theory is more effective in reducing the mass discrepancy between the hydrostatic mass and baryonic mass in GCs. Nevertheless, neither EiBI theory nor BHG completely addresses the mass discrepancy problem. 

\section*{Acknowledgement}
We are sincerely grateful to Prof. Anto Sulaksono and Idrus Husin Belfaqih for precious discussions and comments. We sincerely thank the reviewers for their thorough evaluation of our manuscript. Their insightful comments and constructive suggestions have significantly improved the clarity, accuracy, and overall quality of our work. We deeply appreciate their time and effort in providing such valuable feedback. MLP acknowledges Indonesia Endowment Fund for Education Agency (LPDP) for financial support.





\end{document}